\documentclass[]{spie}  

\usepackage{amsmath,amsfonts,amssymb}
\usepackage{graphicx}
\usepackage{subfigure}
\usepackage[colorlinks=true, allcolors=blue]{hyperref}
\usepackage{svg}
\usepackage{gensymb}
\usepackage{amssymb}
\usepackage{wasysym}
\usepackage{textcomp}
\usepackage{tcolorbox}
\usepackage{caption}
\usepackage{siunitx}

\title{High-Resolution Echelle Spectroscopy for Solar System Planets—A Planet-as-Point-Source Analogy}

\author[a,b]{Parvathy M}
\author[a]{T Sivarani}
\author[a]{S Sriram}
\author[a,b]{Manjunath Bestha}
\author[c]{Devika K Divakar}
\author[a]{S P Rajaguru}
\author[a]{Arun Surya}

\affil[a]{Indian Institute of Astrophysics, II Block Koramangala, Bengaluru-560034, India}
\affil[b]{University of Calcutta, 87/1, College Street, Kolkata-700073, India}
\affil[c]{University of Texas, Austin, TX 78712, United States}

\authorinfo{Further author information: \\ Parvathy Menon: E-mail: parvathy.m@iiap.res.in\\  Sivarani Thirupathi: E-mail: sivarani@iiap.res.in}

\begin{document} 

\maketitle

\begin{abstract}
Transmission spectroscopy has proven to be an effective technique for characterizing exoplanet atmospheres. However, transmission spectroscopy requires planetary transits, which occur for only a small fraction of planetary systems due to geometric alignment constraints; hence, characterizing exoplanets through their reflected spectrum of host stars will be helpful for a large number of exoplanets. The upcoming extremely large telescopes (ELTs) will be able to study the reflected spectra of exoplanets. Here, we present a preliminary optical design and a detailed throughput analysis of the instrumentation that interfaces the 2.34 m Vainu Bappu Telescope prime focus to an existing high-resolution echelle spectrograph with disk-integrated light from solar system objects. One of the primary objectives is to obtain high-resolution, high signal-to-noise reflected spectra from the solar system objects.

\end{abstract}

\keywords{solar system instrumentation, solar system planets, Planet-as-point-source, exoplanet instrumentation, high-resolution spectroscopy, Vainu Bappu Observatory (VBT) }

\section{INTRODUCTION}
\label{sec:intro}  

The discovery of exoplanets in recent decades has transformed planetary science into a fast-expanding, interdisciplinary field that brings together astronomy, atmospheric physics, planetary geology, and astrobiology. Thousands of exoplanets have now been identified, with remarkable diversity in their sizes, orbital characteristics, and host star properties. In spite of the plethora of exoplanets discovered in the past three decades, we still know far less about these exoplanetary systems than we do about the planets in the solar system. Most exoplanets appear only as unresolved point sources, and the spectra we obtain are disk-averaged, low-resolution, and often influenced by instrumental and geometric biases. Characterizing these exoplanet data is still technically challenging and prone to systematic errors. In contrast, the planets within our own solar system offer a unique opportunity to study them in unprecedented detail. So the solar system planets can be used as a testbeds for validating the exoplanet characterization techniques. However, because of the relative distance of the exoplanets, we observe them as point sources, whereas solar system planets extend arcsecond angular sizes on the sky. As a result, the observed spectra of exoplanets cover the entirety of the visible surface, whereas only small portions of a solar system planet can be observed at once with a traditional slit-based spectrograph. Spectra of the solar system planets, in contrast to the disk-averaged spectra of exoplanets, vary according to the local physics and chemistry at the observed location \cite{sagan1993search} and cannot be directly comparable to the exoplanet spectra.\\
Numerous efforts have been made in the past that attempted to observe solar system objects as point sources. \cite{karkoschka1994spectrophotometry, karkoschka1998methane, cowan2009alien, mayorga2016jupiter} The Planet as Exoplanet Analog Spectrograph (PEAS) is a dedicated instrument proposed to observe solar system planets as though they are exoplanets. The instrument achieved first light in 2020 during its initial observation of Mars. \cite{martin2020planet} It is a 0.51 m Ritchey--Chr\'etienn (RC) telescope that directs the light to an integrating sphere, which subsequently transmits disk-integrated light from planets through an optical fiber to an off-the-shelf low-resolution ($  R\approx 500$) spectrograph. Later, the integrating sphere in PEAS has been identified as a considerable factor contributing to light loss, which, in turn, has a detrimental impact on overall throughput. \cite{mannings2022throughput}\\
In this work, we focus on enabling our in-house 2.34 m Vainu Bappu Telescope (VBT) equipped with a fiber-fed High-Resolution Echelle Spectrograph (HiRES) for planet-as-point-source observations. These observations can later serve as a template for future exoplanet observations using ELTs. 


\section{VBT and HiRES Overview}
\label{sec:telescope_overview}

The VBT \cite{bhattacharyya1992vainu} is a 2.34 m equatorial telescope located at the Vainu Bappu Observatory, Kavalur, Tamil Nadu, India. It is a reflecting telescope with an f/3.25 parabolic primary mirror and an f/13 hyperbolic secondary mirror. The secondary mirror can be removed to feed light directly to HiRES at the prime focus. The telescope provides an image scale of 27 arcseconds/mm at prime focus and has a focal length of 7605 mm. VBT is equipped with a fiber-fed High-Resolution Echelle Spectrograph (HiRES) at its prime focus and a low-resolution OMR spectrograph at its Cassegrain focus, making the VBT well-suited for a wide range of dedicated spectral observations across different resolutions.

\begin{table}[h]  

    \centering 
    \begin{tabular}{p{6cm}p{6cm}} 
   
    \hline  \
    \centerline {Telescope and Spectrograph Specifications \cite{rao2005high}} \\ 
    \hline
    \hline
     Primary mirror diameter & ${\diameter}$2.34 m \\
     Primary f/ratio & 3.25\\
     Focal length & 7605 mm\\
     
     Image scale & 27 arcseconds/mm\\
     Size of the fiber core& 100 $\micro$m (2.7 arcseconds) \\
     Resolving power & 27,000, 72,000, 100,000\\
     Wavelength coverage & 400 to 1000 nm \\
     
    \hline 
    \end{tabular}
    \caption{VBT and HiRES Specifications} 
    \label{`vbttable}
\end{table}

    

The light from the prime focus of the telescope is fed to the HiRES using a 100 $\micro$m optical fiber core. HiRES provides spectra with resolving powers of R $\approx$ 27,000 with the full slit open and higher resolutions of R $\approx$ 72,000 and R $\approx$ 100,000 with slit widths of 60 $\micro$m and 30 $\micro$m respectively. Atmospheric seeing at the VBT site typically varies between 1.5 arcseconds and 3.5 arcseconds, with an average seeing of approximately 1.85 arcseconds \cite{sreekanth2019measurements}. VBT and HiRES specifications are given in Table \ref{`vbttable}.

\section{Design of the Instrument Interface Connecting the 2.34 m VBT Prime Focus to the HiRES for Disk-Integrated Observations of Solar System Objects }
\label{sec:design}

\subsection{Requirements}
\label{sec:requirements}
 \begin{figure}[h!]
    \centering
    \subfigure{\includegraphics[trim={3.8cm 4cm 0cm 2.5cm},clip,width=1\textwidth]{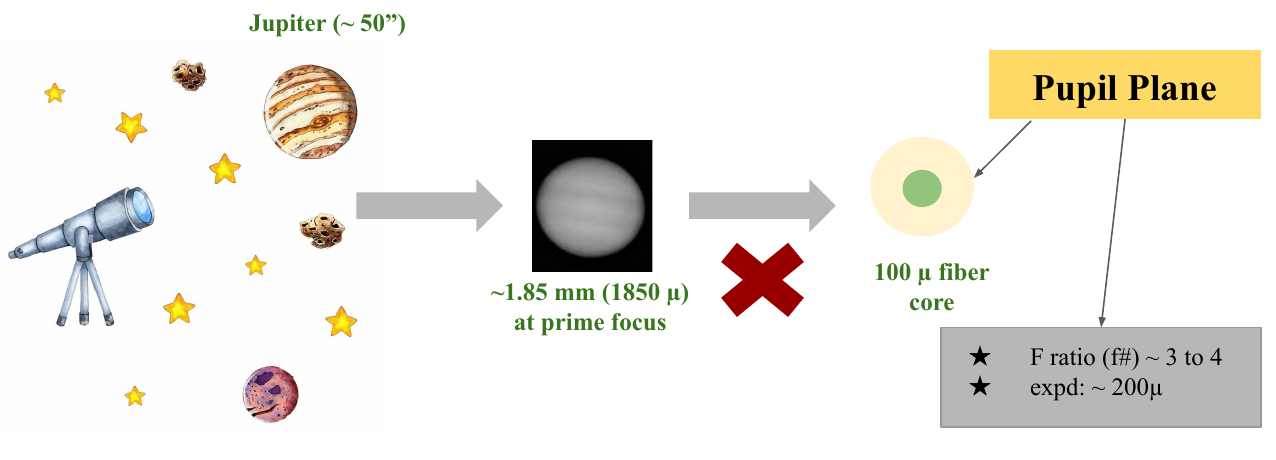}}
    \caption{{Pictorial representation of the requirement for feeding a planet-as-a-point-source from VBT prime focus (a case study with Jupiter)}}
    \label{fig:block_dia}
    
\end{figure} 

The primary objective of this study is to enable the in-house 2.34 m VBT, equipped with the HiRES, for observations of planets treated as point sources. As a case study, let us consider Jupiter, the largest planet in the solar system. Its angular diameter, as observed from Earth, varies between approximately 30 and 50 arcseconds, depending on its distance at the time of observation\cite{nasa-jupiter}. Figure \ref{fig:block_dia} shows the pictorial representation of the requirement for feeding a planet-as-a-point-source from VBT prime focus (a case study with Jupiter).\\
Given the plate scale of 27 arcseconds/mm at the prime focus of the VBT (refer to section \ref{sec:telescope_overview}), a 50-arcsecond-wide object would span roughly 1.85 mm on the focal plane. However, the fiber core used at the VBT prime focus for feeding light into the spectrograph is only 0.1 mm in diameter, making it infeasible to capture the full extent of Jupiter's disk directly. While addressing this limitation, our primary consideration was to avoid making any modifications to the existing telescope or spectrograph optics. Within this constraint, to address this limitation, our approach under consideration is the integration of light at the reimaged pupil plane and sampling a portion of it to the optical fiber. This approach allows us to work within the current optical configuration of the telescope and spectrograph while still enabling the required observations. \\
This method requires the system to achieve an f-ratio of approximately 3 and to produce a pupil image with a diameter of around 200 microns, suitable for coupling into the fiber. The choice of f/3 ensures that the resulting beam is compatible with the numerical aperture of the fiber and thus minimizes optical losses. In addition, to ensure that the spectra are photon-noise limited and suitable for high-resolution analysis, a signal-to-noise ratio (SNR) of at least 50 per resolution element is required. \cite{vbt-echelle-sireesha2019} Since the solar system planets are bright enough, we expect the required SNR of $\geq$ 50 per pixel to be achievable with this approach.

\subsection{Optical Design}
\label{Preliminary Design}
To ensure ease of implementation and cost-effectiveness, we started the design by employing readily available, off-the-shelf optical components. Based on the design requirements, the reimaging optics needed to (i) efficiently couple the pupil onto the fiber input, (ii) maintain a high transmission ($>$ 90\% across 400–700 nm), and (iii) match the numerical aperture (NA $\approx$ 0.22) of the fiber.

\begin{center}
\begin{tabular}{|l|c|c|}
\hline
\textbf{Parameter} & \textbf{Ball Lens} & \textbf{Aspheric Lens} \\
\hline
Diameter & 5 mm & 5 mm \\
Material & LASFN-35 & LB2000 \\
Wavelength & 400 - 2400 nm & 200 - 2000 nm \\
\hline
\end{tabular}
\captionof{table}{Specifications of ball lens and aspheric lens}
\label{tab:edmund_optics_lenses}
\end{center}
A compact microlens arrangement was identified to be sufficient to reimage the pupil onto the fiber input\cite{ansys-pupil}.

\begin{figure}[h!]
    \centering
    \includegraphics[width=0.7\linewidth]{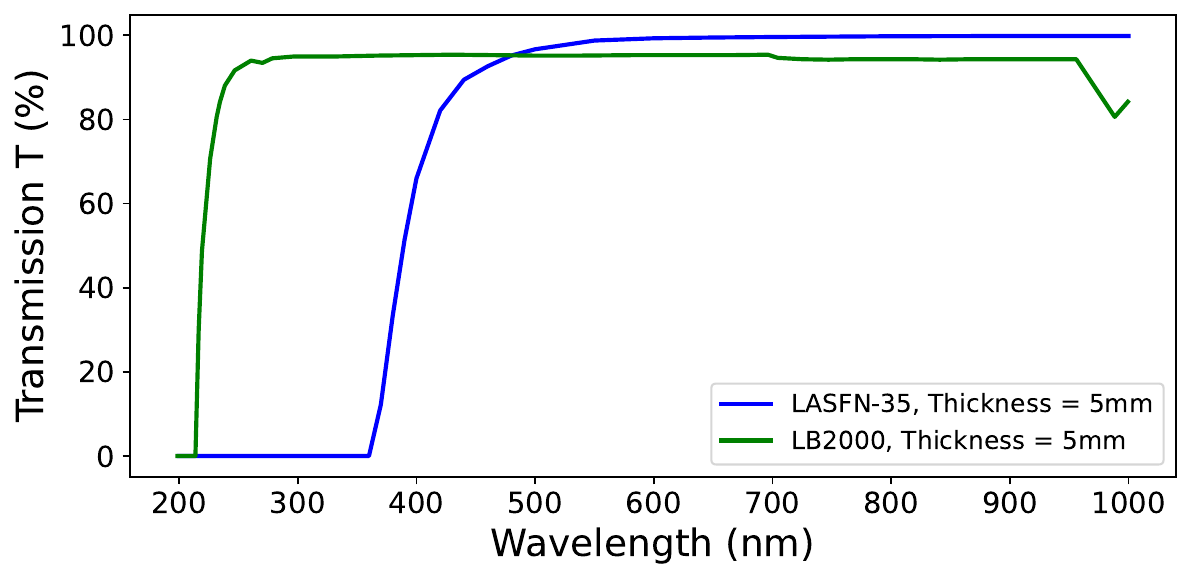}
    \caption{Transmittance of the lens materials LASFN-35 (green) and LB2000 (blue)}
    \label{fig:ltransmission_comparison}
\end{figure}

 We explored the feasibility of using either a ball lens \cite{edmundoptics-balllens-2025} or an aspheric lens \cite{edmundoptics-aspheric-condenser-2025} for this purpose. Both options were chosen based on their high transmission over the operational wavelength range of the spectrograph and their off-the-shelf availability. The specifications of these lenses are summarized in the table \ref{tab:edmund_optics_lenses}, and the transmission efficiency of the lens materials is given in the Figure \ref{fig:ltransmission_comparison}.

\begin{figure}[h!]
    \centering
    \subfigure{\includegraphics[trim={1cm 4cm 1cm 3.95cm},clip,width=1\textwidth]{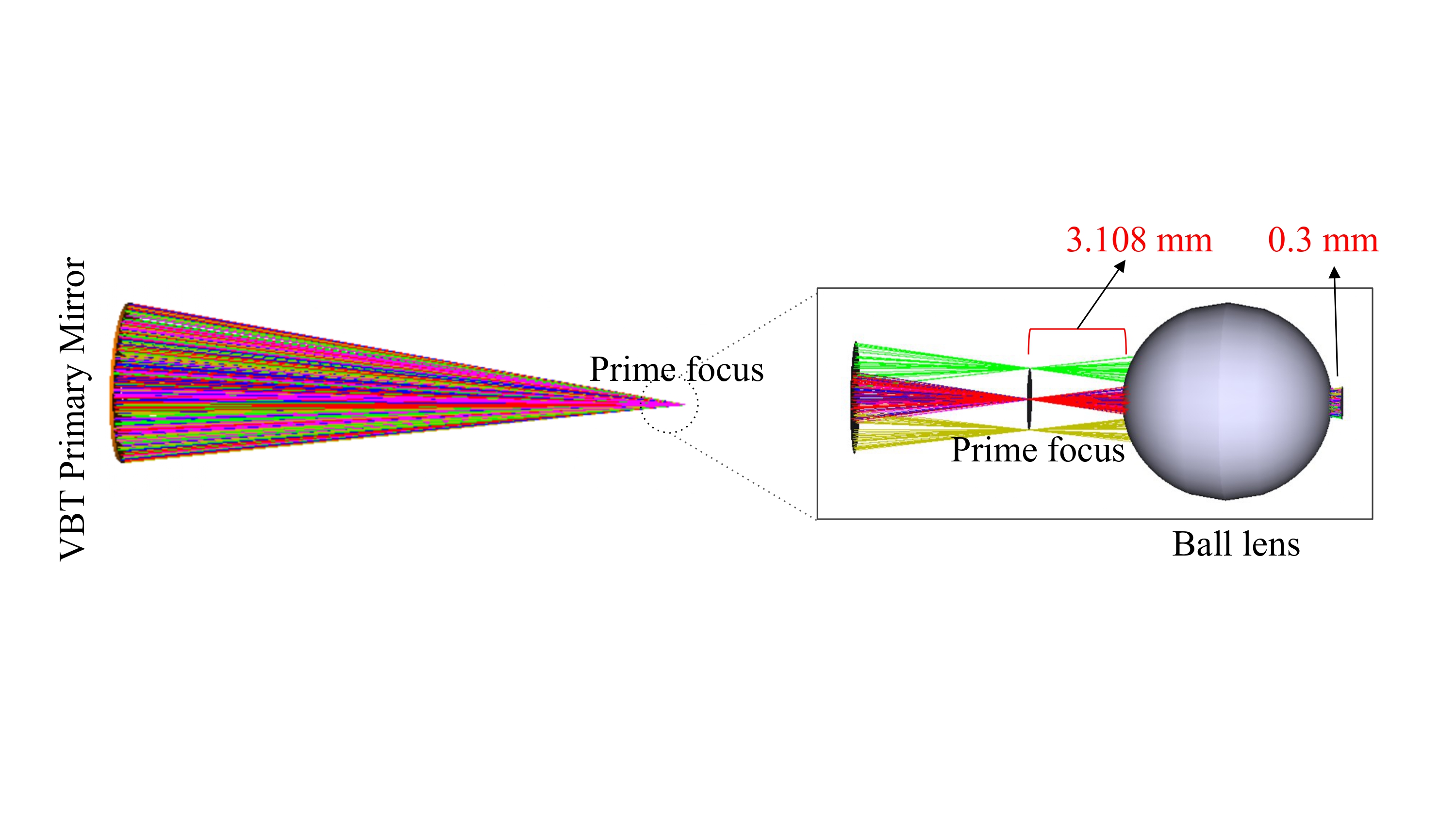}}
    \subfigure{\includegraphics[trim={1cm 4cm 1cm 3cm},clip,width=1\textwidth]{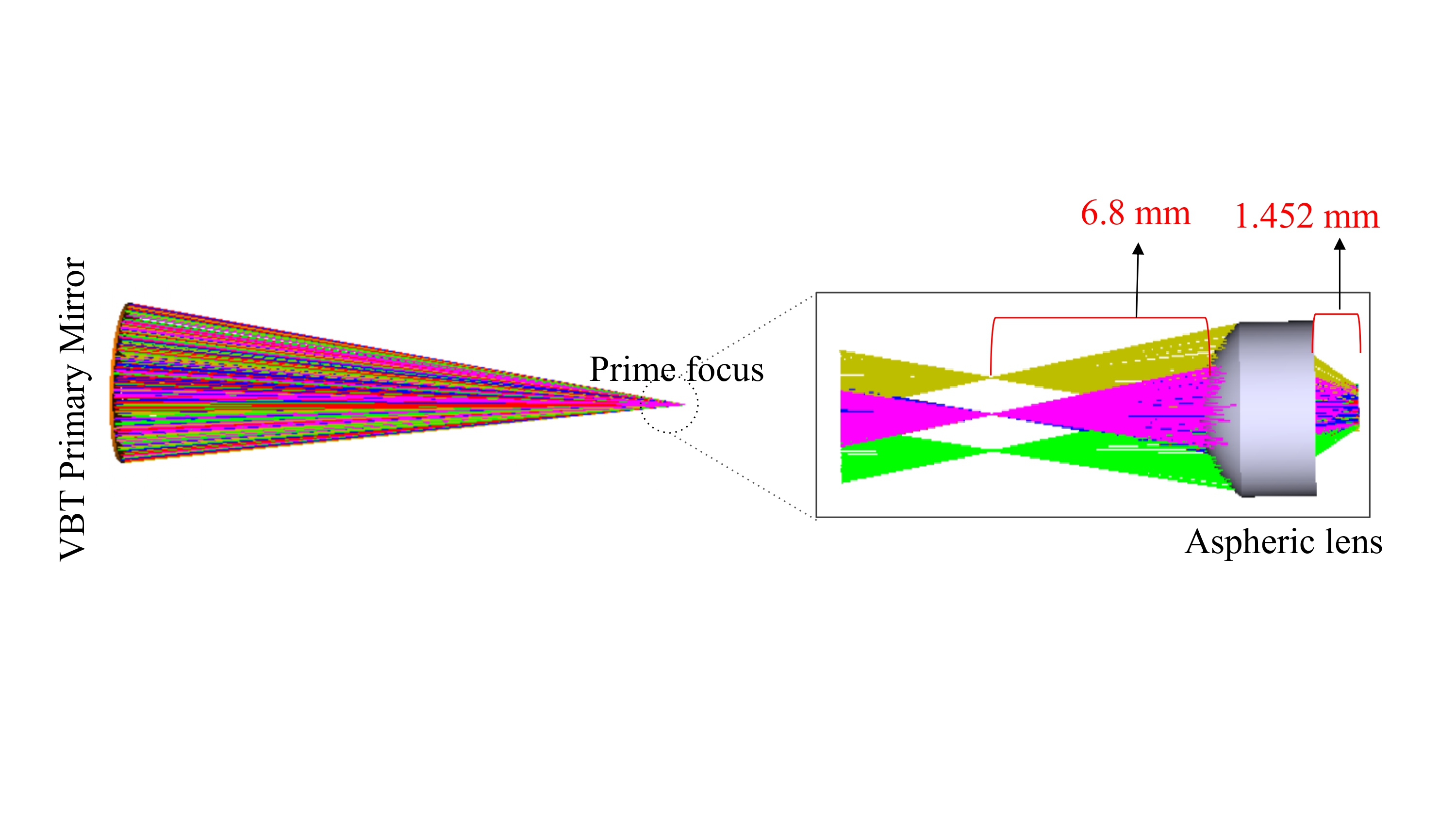}}
    \caption{{Optical layout of two microlens configurations explored for reimaging the VBT prime focus onto the fiber input. (Top) The reimaging optics design with the ball lens. (Bottom) The reimaging optics design with the aspheric lens.}}
    \label{fig:ball_aspheric_optical_layout}
\end{figure}

\begin{figure}[h!]
    \centering
    \subfigure{\includegraphics[trim={1cm 0.6cm 1cm 0.5cm},clip,width=0.65\textwidth]{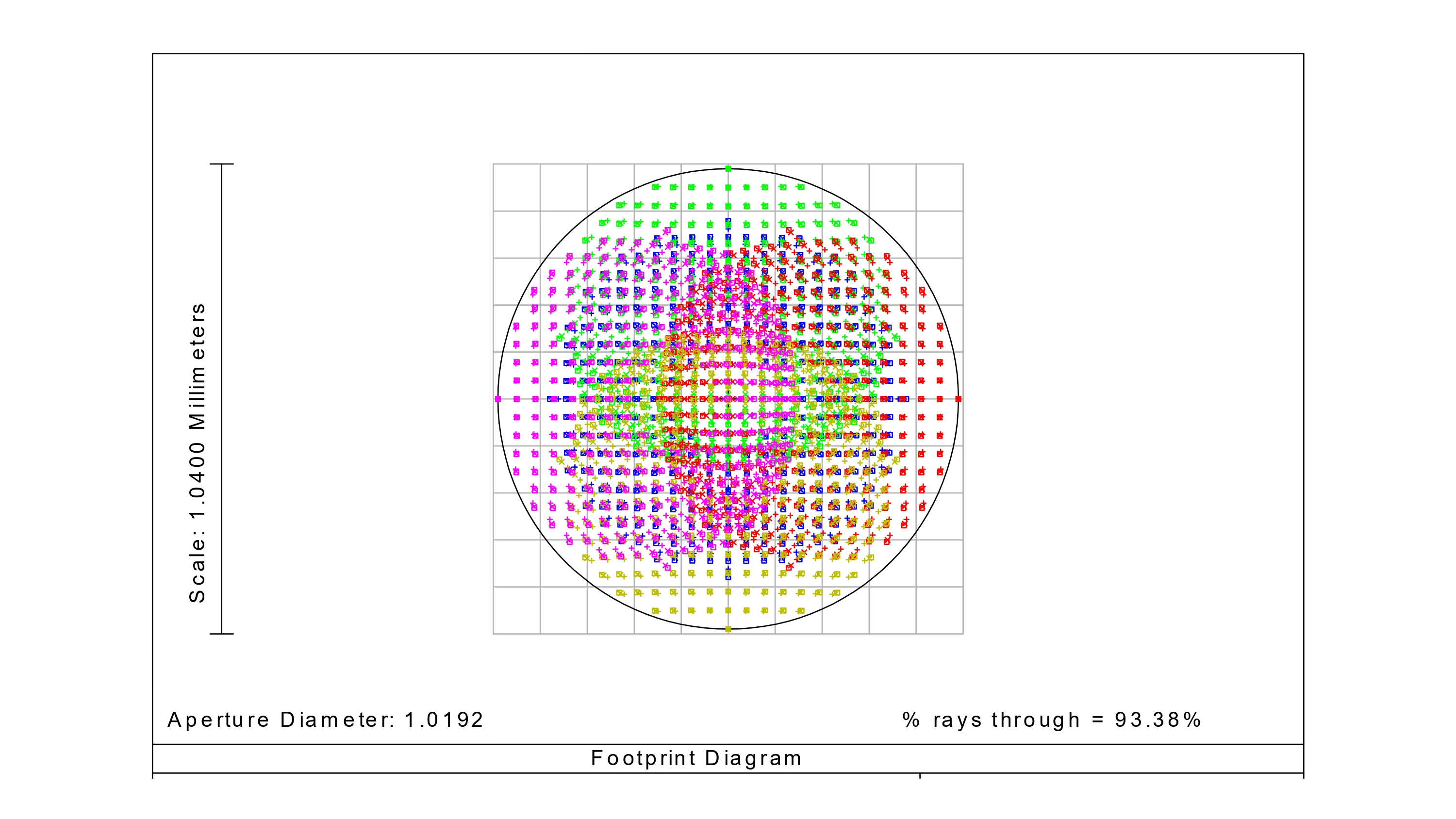}}
    \subfigure{\includegraphics[trim={1cm 0.5cm 1cm 0.5cm},clip,width=0.65\textwidth]{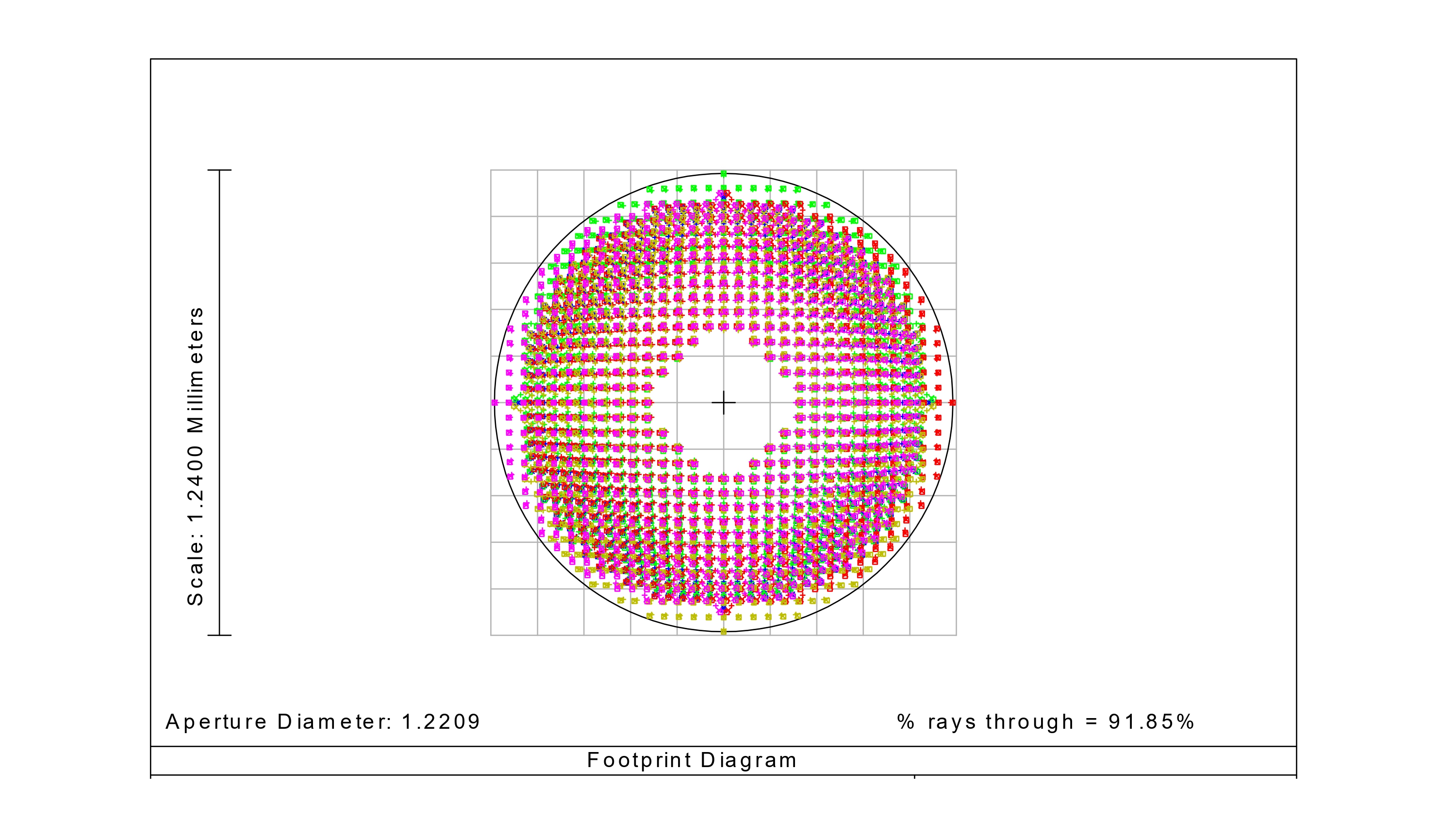}}
    \caption{Footprint diagrams for two microlens configurations used to reimage the VBT prime focus onto the fiber input. The top panel shows the footprint for the ball lens design, and the bottom panel shows the footprint for the aspheric lens design. The aspheric lens configuration produces a more uniform pupil distribution at the fiber input compared to the ball lens}
    \label{fig:ball_aspheric_footprint}
\end{figure}

Ray-tracing simulations were performed using ZEMAX to assess pupil reimaging performance for each case. Both cases delivered the required output f-ratio of approximately 3 while the pupil size is greater than 1 mm. The ball lens offered a compact optical path, with the prime focus located close to the lens surface. The aspheric lens, while larger in physical footprint, produced a more uniform pupil illumination at the fiber input. The optical layout and corresponding pupil footprints for both cases are shown in Figure \ref{fig:ball_aspheric_optical_layout} and Figure \ref{fig:ball_aspheric_footprint}, respectively.

Although larger than the required pupil diameter, the pupil illumination pattern was found to be nearly uniform in the case of the aspheric lens, allowing the fiber to be decentered to collect a portion of the light without significant vignetting. Given the relatively high brightness of solar system objects, it is possible to collect sufficient photons even when only a portion of the pupil is sampled.

\section{Throughput estimates}
\label{Throughput estimates}

The total system throughput of the instrument was estimated by accounting for all major optical and instrumental transmission losses from the telescope aperture to the detector. The calculation was performed at each wavelength using corresponding efficiency curves of individual optical elements and the combined efficiency of the spectrograph\cite{bestha2023atmospheric}. 

The photon flux at the detector per pixel, \(C(\lambda)\), was derived from the planet magnitude and the effective collecting area of the telescope, scaled by the total wavelength-dependent system transmission, as given in equation \ref{eq:count}:\\
\begin{equation}
C(\lambda) = F \times A_{\mathrm{eff}} \times T_{\mathrm{sys}}(\lambda) \times \frac{1}{R * PSF\,Width}
\label{eq:count}
\end{equation}

where F is the photon flux per pixel for the planet, $A_{\mathrm{eff}}$ is the effective telescope area after central obscuration, $T_{\mathrm{sys}}(\lambda)$ represents the cumulative transmission of all optical elements, and R is the resolving power. 

\begin{figure}[h!]
    \centering
    \subfigure[]{\includegraphics[trim={0cm 0cm 0cm 0cm},clip,width=0.65\textwidth]{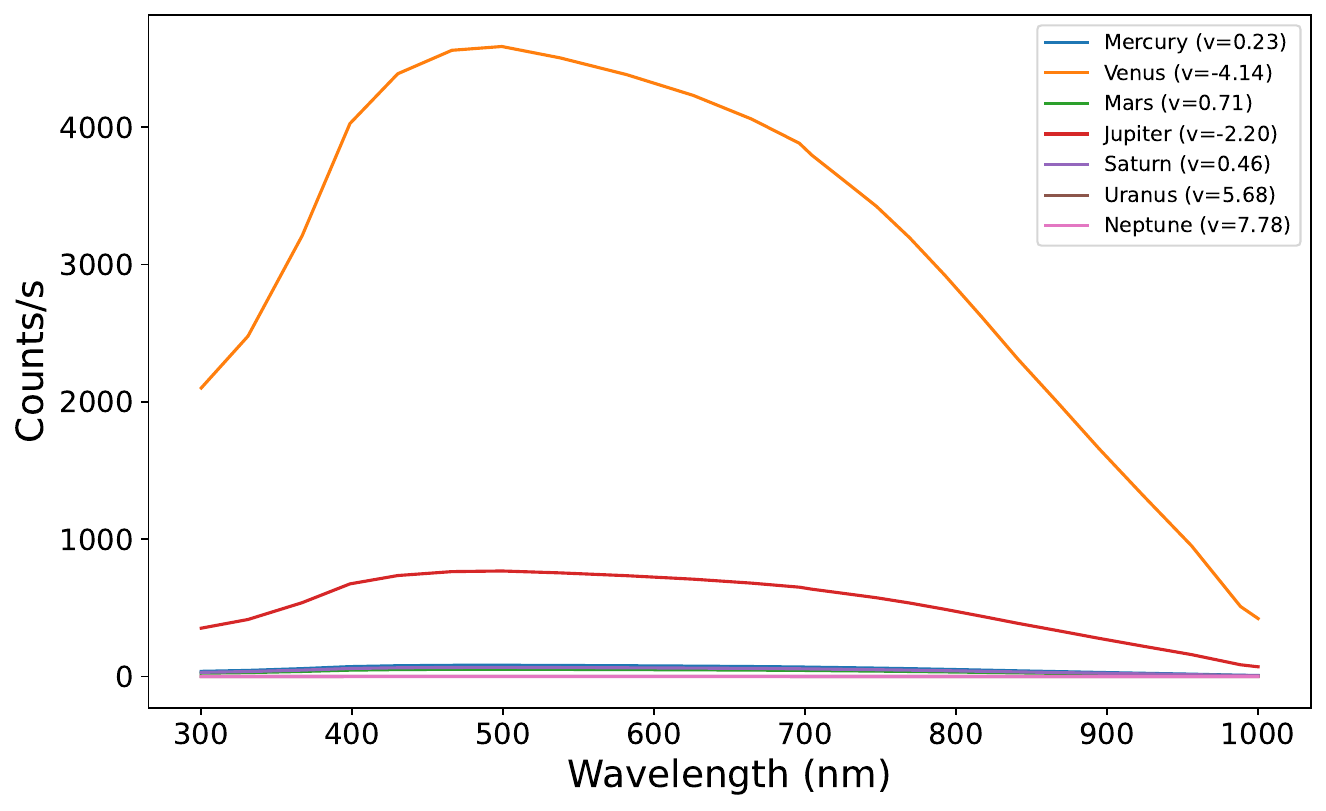}}
    \hfill
    \subfigure[]{\includegraphics[trim={0cm 0cm 0cm 0cm},clip,width=0.65\textwidth]{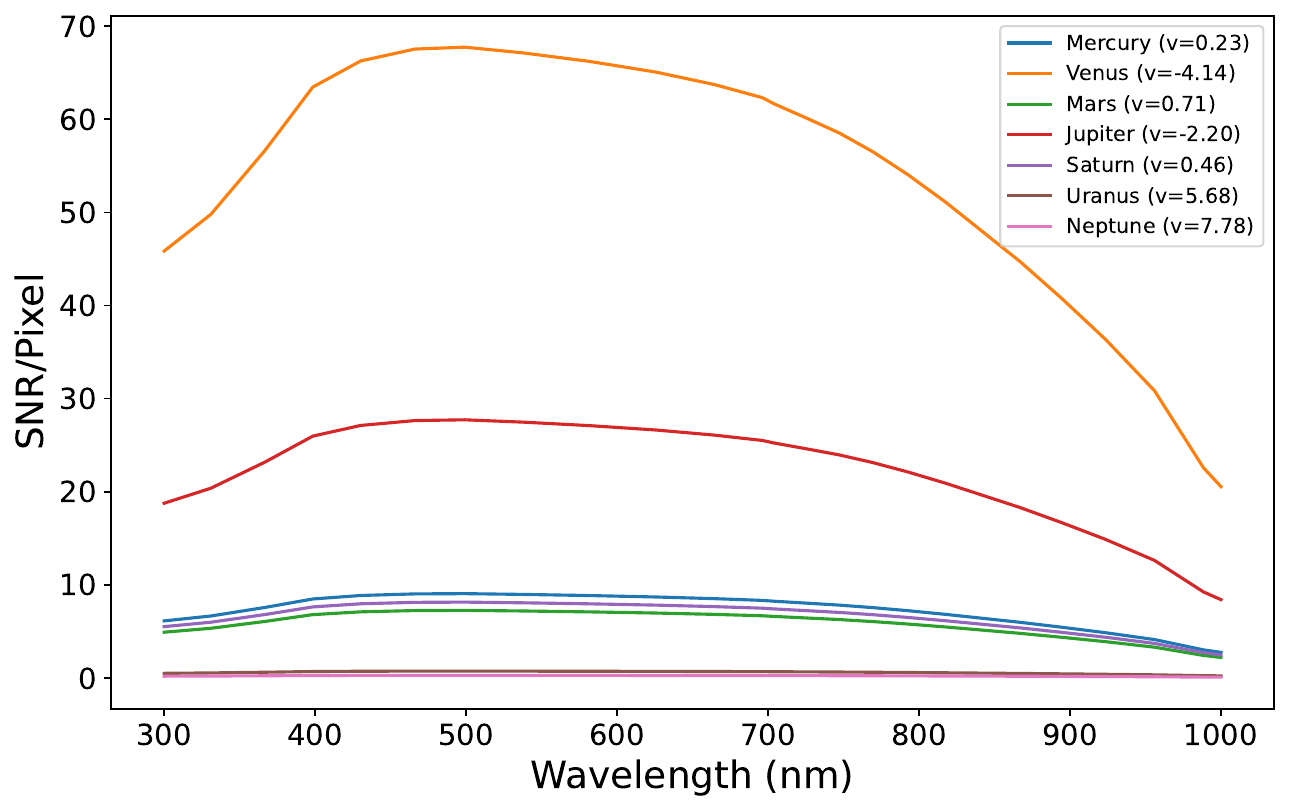}}
    \caption{(a) Calculated photon counts per pixel for each planet across the observed wavelength range. (b) Corresponding signal-to-noise ratio (SNR) per pixel for a 1 s exposure, derived from the total system response.}
    \label{fig:ounts_final}
\end{figure}

The cumulative transmission of all optical elements considered in the calculation included the reflectivity of the primary mirror (50\%), lens transmissivity, area ratio at the pupil plane, Numerical Aperture (NA) ratio, Fresnel loss (8\%), cladding loss (15–17\%)\cite{fti-transmission-loss}, transmissivity over the length of the fiber, spectrograph efficiency, and the detector quantum efficiency (QE). For the detector response, we adopted the quantum efficiency curve of a similar CCD and scaled this to 60\% of the nominal QE values to account for possible degradation over long-term operation. The area ratio at the pupil plane accounts for the loss of light by coupling a pupil portion to the 100 \si{\micro\meter} fiber core. For a cautious estimate of the signal-to-noise ratio (SNR), all transmission efficiencies were considered at their lower bounds.\\
The calculations were performed for all solar system planets using their apparent magnitudes: Mercury (0.23), Venus (–4.14), Mars (0.71), Jupiter (–2.20), Saturn (0.46), Uranus (5.68), and Neptune (7.78) \cite{mallama2018computing}. Figure \ref{fig:ounts_final}a shows the photon counts per pixel for different planets across the observed wavelength range. The corresponding signal-to-noise ratio per pixel is given by the square root of the total number of photons detected per pixel. Figure \ref{fig:ounts_final}b shows the calculated SNR-per-pixel for a 1 s exposure for different planets, illustrating the wavelength dependence driven by the combined efficiency of the system components. The large differences between the planet's counts and SNR arise primarily from differences in their apparent magnitudes, with bright targets such as Venus and Jupiter yielding significantly higher photon flux and SNR compared to the fainter outer planets like Uranus and Neptune. The SNR for the fainter planets can in principle be improved by increasing the exposure time, though other noise source may become limiting.

\section{Mechanical Feasibility}
\label{Mechanical Feasibility}

The next step was to investigate the mechanical feasibility of incorporating additional optics at the prime focus of the VBT. The mechanical constraints were examined by referring to the existing autoguider layout, as described in C Sireesha, PhD thesis (2019) \cite{vbt-echelle-sireesha2019}.

\begin{figure}[h!]
    \centering
    \subfigure{\includegraphics[trim={0cm 3.5cm 0cm 2cm},clip,width=\textwidth]{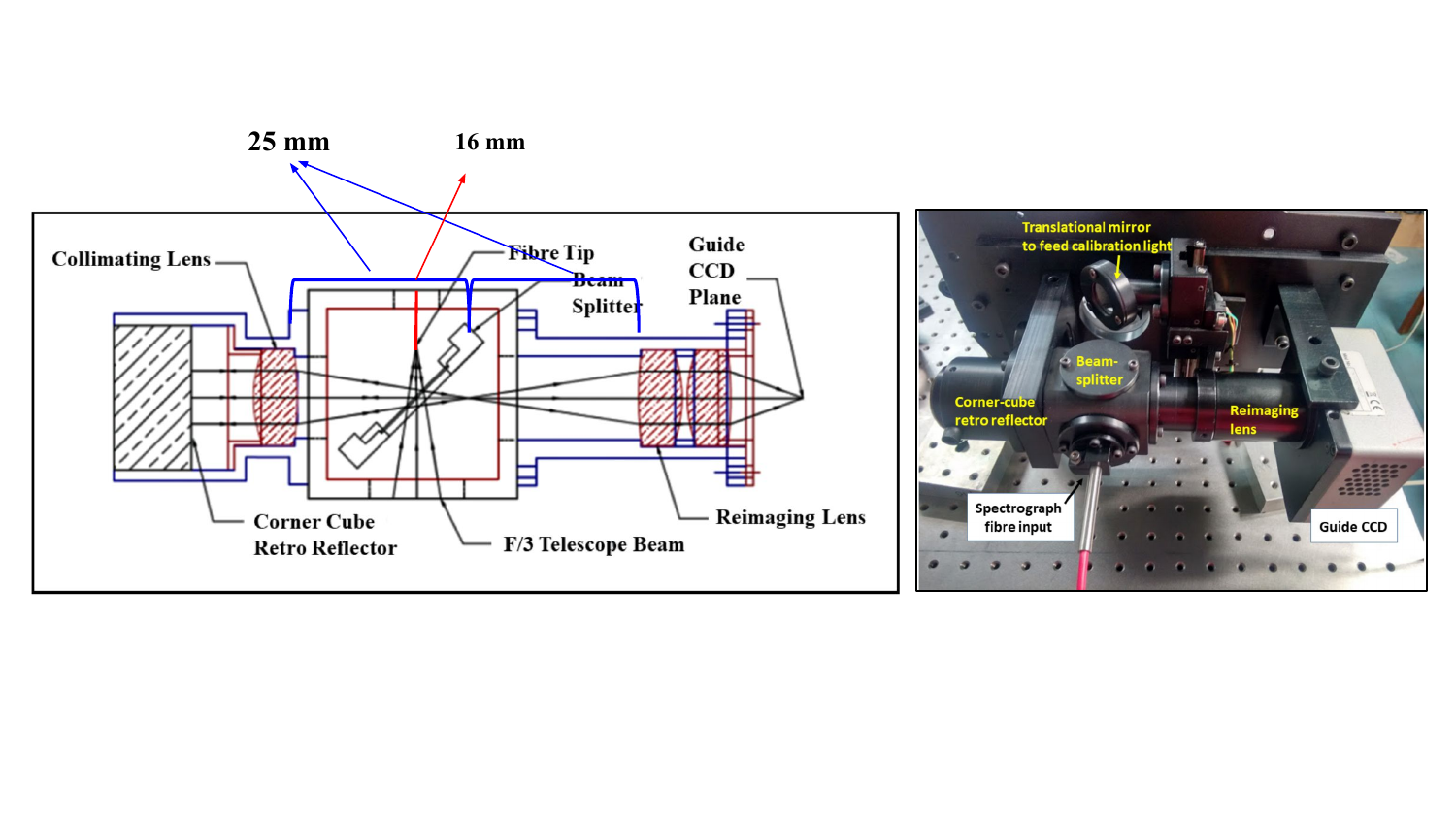}}
    \caption{Optical layout (left) and hardware setup (right) of the VBT autoguider. Starlight from the f/3 primary mirror is split by the beam-splitter, sending 8\% of the light to the guide CCD through a collimating lens, a corner cube retro-reflector, and re-imaging optics for on-axis guiding, while 92\% enters the spectrograph fiber. The guiding algorithm maintains fiber alignment on the CCD. \cite{vbt-echelle-sireesha2019}}
    \label{fig:autoguider}
    
\end{figure}

The schematic diagram on the left of Figure \ref{fig:autoguider} illustrates the current optical configuration of the autoguider mounted at the prime cage of the VBT. Starlight from the f/3 primary mirror falls on the beam-splitter, which reflects 8\% of the beam towards the guide CCD through a re-imaging lens for on-axis target guiding, while the remaining 92\% is directed into the spectrograph fiber. The fiber was re-imaged onto the guiding CCD using a collimating lens, a corner cube retro-reflector, and a re-imaging lens. The fiber position was estimated on the guide CCD, and the guiding algorithm redirects
the starlight onto the fiber coordinates. \cite{vbt-echelle-sireesha2019}\\

The distance from the prime focus to the outer enclosure (or cover) of the autoguider is measured to be approximately 16 mm. In our proposed design, the lens must be placed at a distance of about 7 mm from the prime focus to meet the required criteria. This introduces a mechanical constraint, as it is not feasible to directly place the lens in the current configuration. Shifting the prime focus outside the enclosure will introduce a dimension mismatch. Consequently, two possible solutions are:
\begin{itemize}
    \item{Redesigning the autoguider assembly to create sufficient space for the optical element while maintaining the alignment, or}
    \item{Employing a mechanical ferrule or similar mounting mechanism to insert the lens into the existing setup without significant structural modifications.} 
\end{itemize}
Further analysis is required to identify the most effective method for integrating the lens at the prime focus without affecting the autoguider’s performance.


\section{Conclusion \& Future Work}
\label{conclusion}

This study explored the feasibility of enabling point-source observations of bright solar system objects with the 2.34 m VBT and HiRES without modifying the primary spectrograph optics. Accordingly, the system was designed to reimage the planetary light at the pupil plane and couple it efficiently into a 100 µm fiber core at an F-ratio of approximately 3. The design was demonstrated with an off-the-shelf aspheric microlens offering better performance within the constraints. Throughput estimates confirmed that sufficient photon counts and signal-to-noise ratios can be achieved across the operational wavelength range for bright planetary targets such as Venus and Jupiter, while longer exposures can compensate for fainter outer planets. Mechanically,  the 16 mm spacing between the prime focus and the autoguider enclosure limits placing the lens at the required 7 mm distance, necessitating careful integration strategies for the microlens at the pupil plane.

\section{Acknowledgements}
\label{Acknowledgement}
The authors gratefully acknowledge Prof. Andy Skemer at the University of California, Santa Cruz (UCSC), and the Other Worlds Laboratory (OWL) Exoplanet Summer Program 2024 for invaluable discussions and insights. We thank the technical staff at the Indian Institute of Astrophysics and Vainu Bappu Observatory for their support with the telescope and spectrograph operations. 

\bibliography{report} 

\begin{thebibliography}{10}

\bibitem{sagan1993search}
Sagan, C., Thompson, W.~R., Carlson, R., Gurnett, D., and Hord, C., ``A search for life on earth from the galileo spacecraft,'' {\em Nature}~{\bf 365}(6448),  715--721 (1993).

\bibitem{karkoschka1994spectrophotometry}
Karkoschka, E., ``Spectrophotometry of the jovian planets and titan at 300-to 1000-nm wavelength: The methane spectrum,'' {\em Icarus}~{\bf 111}(1),  174--192 (1994).

\bibitem{karkoschka1998methane}
Karkoschka, E., ``Methane, ammonia, and temperature measurements of the jovian planets and titan from ccd--spectrophotometry,'' {\em Icarus}~{\bf 133}(1),  134--146 (1998).

\bibitem{cowan2009alien}
Cowan, N.~B., Agol, E., Meadows, V.~S., Robinson, T., Livengood, T.~A., Deming, D., Lisse, C.~M., A'Hearn, M.~F., Wellnitz, D.~D., Seager, S., et~al., ``Alien maps of an ocean-bearing world,'' {\em The Astrophysical Journal}~{\bf 700}(2),  915 (2009).

\bibitem{mayorga2016jupiter}
Mayorga, L., Jackiewicz, J., Rages, K., West, R.~A., Knowles, B., Lewis, N., and Marley, M.~S., ``Jupiter’s phase variations from cassini: a testbed for future direct-imaging missions,'' {\em The Astronomical Journal}~{\bf 152}(6),  209 (2016).

\bibitem{martin2020planet}
Martin, E.~C., Skemer, A.~J., Radovan, M.~V., Allen, S.~L., Black, D., Deich, W.~T., Fortney, J.~J., Kruglikov, G., MacDonald, N., Marques, D., et~al., ``The planet as exoplanet analog spectrograph (peas): design and first-light,'' in [{\em Ground-based and Airborne Instrumentation for Astronomy VIII}{\nolinebreak\hspace{0.1em}]},   {\bf 11447},  135--143, SPIE (2020).

\bibitem{mannings2022throughput}
Mannings, A.~G., Martin, E.~C., Skemer, A.~J., Radovan, M.~V., MacDonald, N., Kupke, R., Morris, E.~C., and DeMartino, M., ``Throughput modeling of the planet as exoplanet analog spectrograph (peas),'' in [{\em Ground-based and Airborne Instrumentation for Astronomy IX}{\nolinebreak\hspace{0.1em}]},   {\bf 12184},  1894--1901, SPIE (2022).

\bibitem{bhattacharyya1992vainu}
Bhattacharyya, J. and Rajan, K., ``Vainu bappu telescope,'' {\em Bulletin of the Astronomical Society of India}~{\bf 20},  319--343 (1992).

\bibitem{rao2005high}
Rao, N.~K., Sriram, S., Jayakumar, K., and Gabriel, F. {\em Journal of Astrophysics and Astronomy}~{\bf 26}(2),  331--338 (2005).

\bibitem{sreekanth2019measurements}
Sreekanth, R.~V., Banyal, R.~K., Sridharan, R., and Selvaraj, A., ``Measurements of atmospheric turbulence parameters at vainu bappu observatory using short-exposure ccd images,'' {\em Research in Astronomy and Astrophysics}~{\bf 19}(5),  074 (2019).

\bibitem{nasa-jupiter}
{NASA Planetary Fact Sheet}, ``Jupiter fact sheet.'' \url{https://nssdc.gsfc.nasa.gov/planetary/factsheet/jupiterfact.html} (2025).

\bibitem{vbt-echelle-sireesha2019}
Sireesha, C., ``High precision radial velocity studies on vbt echelle spectrograph.'' \url{https://prints.iiap.res.in/handle/2248/8045} (Aug. 2019).

\bibitem{ansys-pupil}
{Ansys Optics Knowledgebase}, ``Paraxial vs. real pupils in optical system.'' \url{https://optics.ansys.com/hc/en-us/articles/42661958583571-Paraxial-vs-Real-pupils-in-optical-system} (2023).

\bibitem{edmundoptics-balllens-2025}
{Edmund Optics}, ``5.0 mm diameter, lasfn-35 ball lens.'' \url{https://www.edmundoptics.in/p/50mm-diameter-lasfn-35-ball-lens/7168/} (2025).

\bibitem{edmundoptics-aspheric-condenser-2025}
{Edmund Optics}, ``5mm dia. x 4mm fl, uncoated molded aspheric condenser lens.'' \url{https://www.edmundoptics.in/p/5mm-dia-x-4mm-fl-uncoated-molded-aspheric-condenser-lens/34726/} (2025).

\bibitem{bestha2023atmospheric}
Bestha, M., Hasan, A., Divakar, D., Surya, A., Sriram, S., Sivarani, T., Prakash, A., Yadav, S., et~al., ``Atmospheric dispersion corrector for a multi-object spectroscopic mode of hros-tmt,'' in [{\em Astronomical Optics: Design, Manufacture, and Test of Space and Ground Systems IV}{\nolinebreak\hspace{0.1em}]},   {\bf 12677},  267--273, SPIE (2023).

\bibitem{fti-transmission-loss}
{Fiberoptics Technology Inc.}, ``Transmission loss.'' \url{https://www.fiberopticstech.com/technical/transmission-loss/}.

\bibitem{mallama2018computing}
Mallama, A. and Hilton, J.~L., ``Computing apparent planetary magnitudes for the astronomical almanac,'' {\em Astronomy and computing}~{\bf 25},  10--24 (2018).

\end{thebibliography}

\bibliographystyle{spiebib} 
\end{document}